\documentclass[aps,prb,twocolumn,showpacs,superscriptaddress]{revtex4}

\usepackage{graphicx}

\begin{document}

\title
{Monitoring of band gap and magnetic state of graphene nanoribbons
through vacancies}

\author{M. Topsakal}
\affiliation{UNAM-Institute of Materials Science and
Nanotechnology, Bilkent University, Ankara 06800, Turkey}
\author{E. Akt\"{u}rk}
\affiliation{UNAM-Institute of Materials Science and
Nanotechnology, Bilkent University, Ankara 06800, Turkey}
\author{H. Sevin\c{c}li}
\affiliation{UNAM-Institute of Materials Science and
Nanotechnology, Bilkent University, Ankara 06800, Turkey}
\affiliation{Department of Physics, Bilkent University, Ankara
06800, Turkey}
\author{S. Ciraci} \email{ciraci@fen.bilkent.edu.tr}
\affiliation{UNAM-Institute of Materials Science and
Nanotechnology, Bilkent University, Ankara 06800, Turkey}
\affiliation{Department of Physics, Bilkent University, Ankara
06800, Turkey}


\begin{abstract}
Using first-principles plane wave calculations we predict that
electronic and magnetic properties of graphene nanoribbons can be
affected by defect-induced itinerant states. The band gaps of
armchair nanoribbons can be modified by hydrogen saturated holes.
Defects due to periodically repeating vacancy or divacancies
induce metallization, as well as magnetization in non-magnetic
semiconducting nanoribbons due to the spin-polarization of local
defect states. Antiferromagnetic ground state of semiconducting
zigzag ribbons can change to ferrimagnetic state upon creation of
vacancy defects, which reconstruct and interact with edge states.
Even more remarkable is that all these effects of vacancy defects
are found to depend on their geometry and position relative to
edges. It is shown that these effects can, in fact, be realized
without really creating defects \end{abstract}

\pacs{73.22.-f, 75.75.+a, 75.70.Ak}

\maketitle

\section{introduction}

Its unusual electronic energy band structure and charge carriers
resembling massless Dirac Fermions have made graphene honeycomb
structure an active field of
research\cite{novo,zhang,berger,kats,geim}. Quasi 1D graphene
ribbons have even more interesting electronic and magnetic
properties depending on their size and symmetry\cite{ribbon}.
These are edge states of zigzag ribbons with opposite spin
polarization\cite{fujita} and band gaps varying with the width of
the ribbon\cite{cohen}.

Theoretical studies have predicted that energetic electrons and ions
can induce polymorphic atomic defects , such as vacancies in graphene
\cite{nordlund}. Using high-resolution TEM the observation of vacancies 
have been reported \cite{lijima}.
Recent studies have shown that vacancies created on
two-dimensional (2D) graphene by high-energy electron or ion
irradiation can induce magnetism in a system consisting of only
$sp$-electrons\cite{esquinazi,yazyev}. It has been argued that
Stoner magnetism with high $T_C$ originates from the
spin-polarized extended states induced by the vacancy defects,
while RKKY coupling is suppressed. These effects of defects on 1D
semiconducting graphene nanoribbons should be more complex and
interesting, because their band gap, magnetic state and symmetry
are expected to intervene.

Present study based on extensive first-principles as well
as empirical tight-binding calculations (ETB) has shown
that the band gap and magnetic state of any armchair or zigzag
nanoribbons can be modified by single or multiple vacancies
(holes). The effects of these defects depend on their symmetry,
repeating periodicity and positions. Even if the formation of
periodic defects of desired symmetry may not easily be achieved,
similar effects can be created through the potential difference
across a ribbon applied by periodically arranged tips. When
combined with various properties of nanoribbons these results can
initiate a number of interesting applications. We believe that our
results are important for further studies, since the graphene
ribbons can now be produced with precision having widths sub 10
nm\cite{dai1} and nanodevices can be fabricated
thereof\cite{dai2}.

\section{Model and Methodology}
We have performed first-principles plane wave calculations within Density Functional Theory 
(DFT) using PAW potentials \cite{Bleochl}. The exchange correlation potential has been
approximated by Generalized Gradient Approximation (GGA) using
PW91 \cite{pw91} functional both for spin-polarized and
spin-unpolarized cases. All structures have been treated within
supercell geometry using the periodic boundary conditions. A
plane-wave basis set with kinetic energy cutoff of 500 eV has been
used. In the self-consistent potential and total energy
calculations the Brillouin zone (BZ) is sampled by (1x1x35)
special \textbf{k}-points for ribbons. This sampling is scaled
according to the size of superlattices. All atomic positions and
lattice constants are optimized by using the conjugate gradient
method where total energy and atomic forces are minimized. The
convergence for energy is chosen as 10$^{-5}$ eV between two
steps, and the maximum force allowed on each atom is less than 0.02 eV /\AA.
Numerical plane wave calculations have been performed by using
VASP package \cite{vasp1,vasp2}

\section{Vacancy Defects in Nanoribbons}
We start by summarizing the electronic properties of graphene nanoribbons which are relevant for the present study.
We will consider hydrogen terminated nanoribbons if it is not stated otherwise. 
Armchair graphene nanoribbons, AGNR($N$) ($N$ being the number of
carbon atoms in the primitive unit cell), are non-magnetic
semiconductors. The band gap\cite{cohen,gw}, $E_{G}$, of a bare or
hydrogen terminated AGNR($N$) is small for $N=6m-2$ ($m$ being an
integer), but from $N=6m$ to $N=6m+2$ it increases and passing
through a maximum it becomes again small at the next minimum
corresponding to $N=6m+4$. As $E_{G}$ oscillates with $N$ its
value shall decrease eventually to zero as $N \rightarrow \infty$. Bare
and hydrogen terminated zigzag nanoribbons, ZGNR($N$), are also
semiconductors with $E_{G}$ decreasing consistently as $N$
increases for $N>$8, but their edge states give rise to
antiferromagnetic (AFM) ground state\cite{fujita}. We will show that
these magnetic and electronic properties can be modified by
defects originating from vacancies or holes created in those
ribbons.

\begin{figure}
\includegraphics[width=7.95cm]{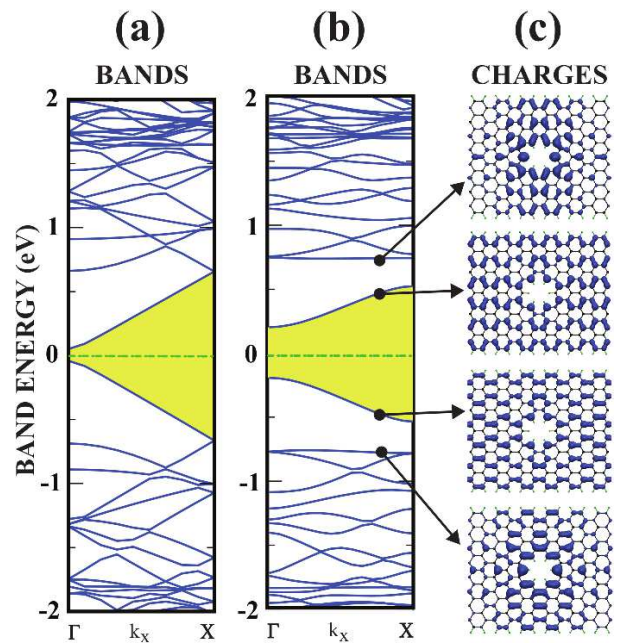}
\caption{(Color online)(a) Energy band structures of AGNR($N$=34;$l$=6)
with and (b) without a hole consisting of
six carbon vacancies. (c) Charge density isosurfaces of selected
states. Carbon atoms (represented by black circles), which have
coordination number lower than 3 are terminated by hydrogen atoms
(represented by small gray circles).} \label{fig:1-a}
\end{figure}

A hole can be created when carbon atoms at the corners of any
hexagon of an armchair ribbon are removed and subsequently
remaining six two-fold coordinated carbon atoms are terminated
with hydrogen atoms. Figure ~\ref{fig:1-a} (a) - (b) shows that the
electronic structure of AGNR($34$) is strongly modified by such a
hole which is placed at the center of the ribbon. The hole repeats itself
at each supercell which comprises six primitive cells
corresponding to a repeat period of $l$=6. This ribbon having a
periodic hole (or defect) is specified as AGNR($N$;$l$) and has
non-magnetic ground state. Despite large separation of periodic
defect which hinders their direct coupling such a strong
modification of the band gap is somehow unexpected. However, it is
an indirect effect and occurs since the itinerant (Bloch) states of band
edges are modified by the defect. At the end, the direct band
gap\cite{gw} at the $\Gamma$-point has widened from 0.09 eV to
0.40 eV due to a defect situated at the center of the ribbon. AGNR (36) exhibit the same behavior; 
namely its band gap increases when a similar hole repeats itself at each supercell consisting of
six primitive cell. However, in contrast, the band gap of  AGNR(38), which is normally larger
than that of  AGNR(34), is reduced if the same hole is
introduced at its center. In addition, states localized around the
defect have formed flat bands near the edge of valence and
conduction bands because of their reduced coupling. 

\begin{figure}
\includegraphics[width=7.95cm]{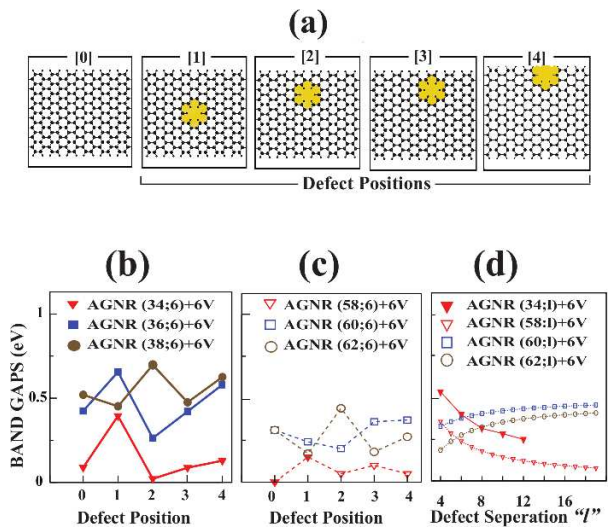}
\caption{(Color online) (a) Positions
of a hole in the ribbon are indicated by numerals. The nanoribbon without a hole is  
specified by ``0''. (b) and (c)
Variations of band gaps of AGNR($N$,$l$) with the position of the
hole specified as 6V . (d) Variation of band gap with repeat
periodicity, $l$. Results for $N < 58$ are obtained by
first-principles calculations.} \label{fig:1-b}
\end{figure}

Moreover, the effect of this defect is strongly
dependent on its position relative to the both edges of the
nanoribbon as depicted in Fig. ~\ref{fig:1-b} (a). As shown in Fig.
~\ref{fig:1-b} (b), the changes in band gap depend on $N$ of AGNR
(in the family specified as $6m+q$, $q$ being -2, 0 and +2) as
well as on the position of the hole. We note that the variation of the band gap
with the position of the hole relative to the edges of the ribbon show similar trends for both AGNR (34,6)
 and AGNR (36;6). However, AGNR (38;6) displays an opposite trend.
 
ETB calculations indicate that a similar behavior is also obtained
in AGNR($N$;$l$)'s having relatively larger $N$ (i.e. $N$=58, 60,
62) in Fig ~\ref{fig:1-b} (c). As expected the effect of the hole on
the band gap depends on the repeat periodicity. As shown in Fig
~\ref{fig:1-b} (d), the effects of defect decrease with increasing
repeat periodicity $l$. We found that larger holes with different
geometry and rotation symmetry can result in diverse electronic
structure and confined states. It should be noted that a repeating
hole can also modify the mechanical properties. For example, the
stiffness of a ribbon is reduced by the presence of a hole. The
force constant, $\kappa=\partial^{2} E_T/\partial c^2$ ($c$ being
the lattice constant) calculated for AGNR(34;6) with a hole at its
center ($\kappa=$ 6.03 eV/\AA) is found to be smaller than that
without a hole $\kappa =$7.50 eV/\AA).

\begin{figure}
\includegraphics[width=7.95cm]{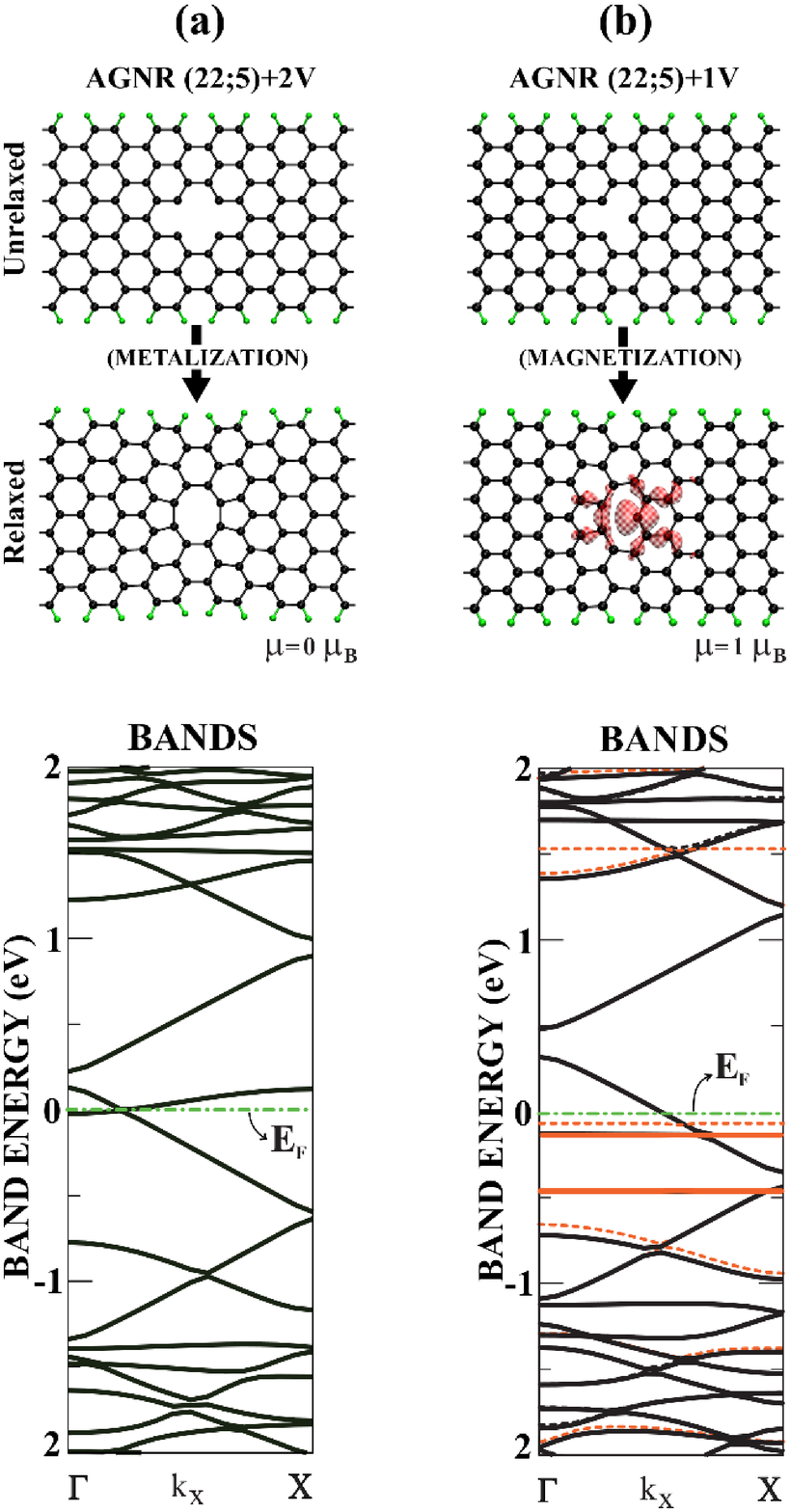}
\caption{(Color online)(a) Metallization of the semiconducting
AGNR(22) by the formation of divacancies with repeat period of
$l$=5. (b) Magnetization of the non-magnetic AGNR(22) by a defect
due to the single carbon atom vacancy with the same repeat
periodicity. Isosurfaces around the vacancy corresponding to $\Delta \rho_{T} $;
namely the difference of the total charge density of different spin directions.Solid (blue) and dashed (red) lines are for spin-up
and spin-down bands; solid (black) lines are nonmagnetic bands.}
\label{fig:2}
\end{figure}

We now show that different types of vacancies in the same armchair 
nanoribbon gives rise to different changes in electronic and magnetic 
properties. A divacancy created in AGNR(22) can cause to a
dramatic change in the electronic state of the ribbon when it is
repeated with the periodicity of $l$=5. Such an armchair ribbon is
specified as AGNR(22;5). The divacancy first relaxes and forms an
eight fold ring of carbon atoms which is adjacent to six hexagons
and two pentagons. In Fig. ~\ref{fig:2} (a) we see that the
non-magnetic and semiconducting AGNR(22) with band gap of
$E_G$=0.18 eV becomes a non-magnetic metal, since a flat band
derived from the defect occurs below the top of the valance band
edge and causes to a metallic state.

The effect of a single carbon vacancy becomes even more
interesting. A single vacancy created in AGNR(22) is relaxed and
the three-fold rotation symmetry is broken due to Jahn-Teller
distortion. At the end, a nine-sided ring forms adjacent to a
pentagon as shown in Fig. ~\ref{fig:2} (b). Owing to the
spin-polarization of the $sp^2$-dangling bond on the two-fold
coordinated carbon atom and adjacent orbitals at the defect site
the system obtains an unbalanced spin. In fact, the difference of
total charge density corresponding to states of different spin
states, i.e. $\Delta \rho_{T} =\rho_T^{\uparrow} -
\rho_T^{\downarrow}$ is non-zero and exhibits a distribution shown
in Fig. ~\ref{fig:2} (b). Because of unbalanced spin, AGNR(22;5)
gains a net magnetic moment of $\mu$=1 $\mu_B$ per cell. This
attributes a magnetic state to the nanoribbon, which was
nonmagnetic otherwise.

Furthermore, the spin degeneracy of some bands related with this
defect is broken and spin-up and spin-down bands are split. The
dispersive, non-magnetic band at the edge of the valence band
becomes partially emptied, since its electrons are transferred to
the flat spin bands below. Eventually, the semiconducting ribbon
becomes metallic.

\begin{figure}
\includegraphics[width=7.8cm]{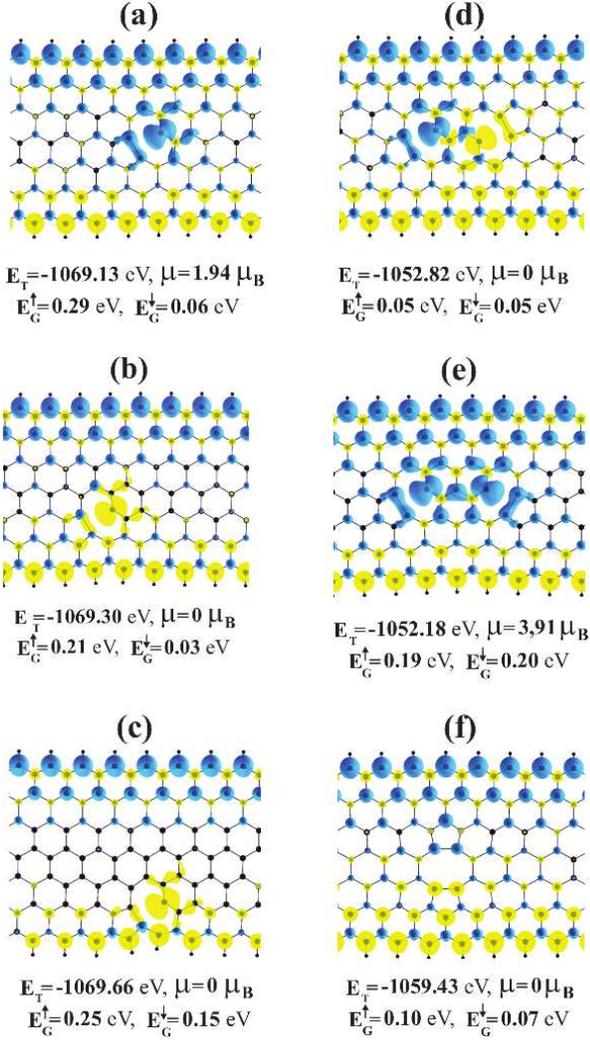}
\caption{(Color online) Vacancy and divacancy formation in an
antiferromagnetic semiconductor ZGNR(14) with repeat period of
$l$=8. Calculated total energy, $E_T$ (in eV/cell), net magnetic
moment, $\mu$ (in Bohr magneton $\mu_B$/cell), band gap between
spin-up(down) conduction and valence bands,
$E_G^{\uparrow(\downarrow)}$) are shown for each case. Blue and yellow isosurfaces corresponds to the difference of spin-up
and spin-down states }
\label{fig:3}
\end{figure}

Not only armchair, but also zigzag nanoribbons are strongly
affected by defects due to single and multiple vacancies. When
coupled with the magnetic edge states of the zigzag nanoribbons
the vacancy defect brings about additional changes.
The magnetic state and energy band structure of these ribbons depend
on the type and geometry of the defects. In Fig. ~\ref{fig:3}
(a)-(c), the effect of a defect generated from the single vacancy
with a repeat periodicity of $l$=8 is examined in ZGNR(14;8) for
three different positions. The total energy is 0.53 eV lowered
when the defect is situated at the edge rather than at the center
of the ribbon.

ZGNR(14;8) has a net magnetic moment of $\mu$=1.94 $\mu_B$ when
the defect is situated at the center of the ribbon and hence its
antiferromagnetic ground state has changed to ferrimagnetic state
through the magnetic moment of the single vacancy. Otherwise,
$\mu$ becomes zero, when the position of the defect deviates from
the center. For example, in Fig. ~\ref{fig:3} (b) and (c) the sum
of the magnetic moments of the edge states is not zero, but the
net magnetic moment per unit cell becomes zero only after the
spins of the defect is added. Even if the net magnetic moment
$\mu$=0, ZGNR(14,8) does not have an antiferromagnetic ground
state due to the presence of single vacancy. The edge states, each
normally having equal but opposite magnetic moments, become
ferrimagnetic when a defect is introduced. The total magnetic
moment of the supercell vanish only after the magnetic moment of
defect has been taken into account. Since the spin-degeneracy has
been broken, one can define $E_G^{\uparrow}$ and
$E_G^{\downarrow}$ for majority and minority spin states. Not only
the magnetic state, but also the band gap of zigzag ribbons in
Fig.~\ref{fig:3} are affected by the symmetry and the position of
the defect relative to edges. In Fig. ~\ref{fig:3} (d)-(f) two
defects associated with two separated vacancy and a defect
associated with a relaxed divacancy exhibit similar behaviors.

An important issue to be addressed here is the breakdown of Lieb's
theorem\cite{lieb} for those zigzag ribbons.  According to Lieb's
theorem, the net magnetic moment per cell is determined with the
difference in the number of atoms belonging to different
sublattices, and it shall be either $\mu$=1 $\mu_B$ or 2 $\mu_B$
for the cases in  Fig.~\ref{fig:3}. None of the cases in Fig.
~\ref{fig:3} is in agreement with Lieb's theorem. Here one can
consider two features, which may be responsible from this
discrepancy. First is the strong Jahn-Teller distortion and
relaxation of carbon atoms at the close proximity of the defect.
As a result some dangling $sp^2$-bonds reconstructed to form new C-C
covalent bonds. The lowering of the total energy, a driving force
for such reconstruction, is as high as 0.5-0.6 eV/cell. Second is
the interaction with the magnetic edge states, which becomes
effective for narrow ZGNR's.

\begin{figure}
\includegraphics[width=7.95cm]{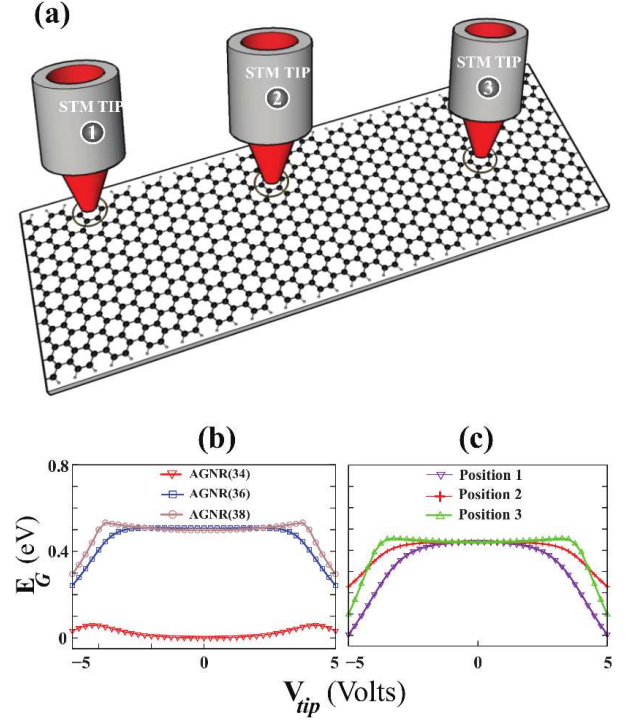}
\caption{(Color online) (a) Monitoring of band gaps $E_G$ by
applying a local bias voltage $V_{tip}$ across the ribbon. (b)
$E_G$ versus $V_{tip}$ applied at the center of AGNR($N$;6) for
$N$=34, 36 and 38. (b) $E_G$ versus $V_{tip}$ of AGNR(34) for
different tip positions schematically described at the top. The
tip (or electrodes) are situated at one of the position 1-3. The
repeat period is $l$=20.}
\label{fig:4}
\end{figure}

Finally, we note that introducing periodic vacancies or defects on
a given ribbon appear to be difficult with the state of the art
technologies. Here we propose a method as described in Fig.
~\ref{fig:4} to achieve the formation of periodic local defects
like holes or vacancies. The sharp electrodes like STM tips are
situated at desired locations, such as one of the cases 1-3 in
Fig. ~\ref{fig:4} (a) on the graphene with a given repeat
periodicity. A potential difference, $V_{tip}$ common to all
electrodes (tips) is applied between the tip and underlying
insulator through the graphene. This way the electronic potential
of graphene atoms just below the tip is locally lowered or raised
depending on the polarity of $V_{tip}$. Here, the effect of
locally and periodically applied potential difference has been
modelled by ETB, where the on-site energies of carbon atoms below
the tip have been changed accordingly. Although the present model
is crude, it still allows us the realization of monitoring of the
properties of nanoribbons. In Fig. ~\ref{fig:4} (b) the variation
of $E_G$ with $V_{tip}$ is calculated for AGNR(N;20) with $N$=34,
36 and 38 by using periodically located tips at the center of the
ribbon. Because of the electron-hole symmetry in the ribbons, the
band gap variation depends on the magnitude of the bias voltage.
In Fig. ~\ref{fig:4} (c) the variation of the band gap of
AGNR(42;20) with $V_{tip}$ and position of the tip is shown. In
spite of the fact that the modifications of the band gaps are not
the same as in Fig. ~\ref{fig:1-b}, the available parameters, such
as $V_{tip}$, $l$, tip-geometry and  its position make the
monitoring of the properties possible.

\section{Conclusions}
In conclusion, we show that the energy band gaps and magnetic
states of graphene nanoribbons can be modified by defects due to
single or multiple vacancies. The optimized atomic structure of 
various vacancy defects have been found to be identical with 
the TEM images reported earlier  \cite{lijima}. Two different electronic states are
distinguished: These are: (i) itinerant Bloch states perturbed by
defects, (ii) defect induced states. While the former is
dispersive, the latter give rise to flat bands. The reconstruction
and spin-polarization of the orbitals at the close proximity of
the defect give rise to net magnetic moments, which, in turn,
changes the magnetic ground state of the defect-free ribbon.

\begin{acknowledgments}
Part of the computations have been carried out by using UYBHM at
Istanbul Technical University through a grant (2-024-2007).
\end{acknowledgments}


\begin{references}
\bibitem{schottky}
Using the definition of Schottky defect, we calculated the
formation energy of relaxed vacancy of single carbon atom in
AGNR(34) to be 7.79 eV. This value is smaller than the binding
energy of a carbon atom in the graphene, which is calculated to be
9.23.
\end{references}
\end{document}